# Addressing the World War 2 Warm Anomaly in HadSST.4.2.0.0

Caroline Sandford* and Nick Rayner

Met Office Hadley Centre, Exeter, EX1 3PB

* caroline.sandford@metoffice.gov.uk

## Funding information

Caroline Sandford and Nick Rayner were supported by the Met Office Hadley Centre Climate Programme funded by DSIT.

## Abstract

We present an update to the Hadley Centre Sea-Surface Temperature dataset (HadSST.4.2.0.0) that addresses residual warm bias during the Second World War (WW2). Using an existing quantitative definition of the WW2 warm anomaly we identify Engine Room Intake (ERI) bias corrections as the dominant factor in this warm bias in HadSST4, and use this to propose new constraints on ERI bias estimates prior to 1950. In addition, we implement corrections for truncation bias in observations from the Japanese Kobe Collection, spanning the period from 1933 to 1961. We evaluate the effects of these changes with respect to the previous version of HadSST and compare with the most recent iterations of other SST datasets including ERSSTv6, COBE-SST3 and DCENT. We show that it is possible to remove the WW2 warm anomaly using a physically-based approach that maintains the independence of HadSST from land surface temperature records, and preserves structural diversity within the range of available global SST datasets.

## Graphical abstract

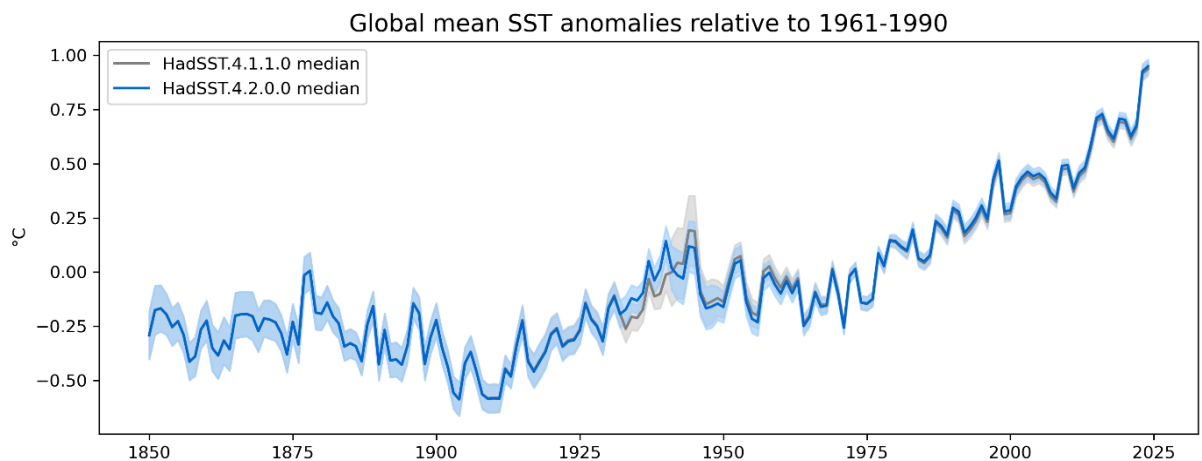

We present an update to the Hadley Centre Sea-Surface Temperature dataset (HadSST.4.2.0.0) that addresses residual warm bias during the Second World War (WW2). Using a quantitative definition of the WW2 warm anomaly we identify Engine Room Intake (ERI) bias corrections as the dominant factor in HadSST4, and use this to propose new constraints on ERI bias estimates prior to 1950. We show that it is possible to remove the WW2 warm anomaly while preserving independence from land surface temperature records.

## 1. Introduction

Given the current state of the global climate, historical temperature records that extend reliably back to the pre-industrial period (1850-1900) are of substantial interest and importance. As oceans make up over 70% of the earth's surface, global near-surface temperature records are strongly influenced by historical reconstructions of sea-surface temperature.

The four temperature datasets contributing to the 6[th] Assessment Report of the Intergovernmental Panel on Climate Change (AR6) Working Group 1 report are each amalgamated from separate land and sea-surface temperature components (Gulev et al. 2021). However, on a global average, historical sea-surface temperature (SST) and marine air temperature records are observed to differ from land near-surface air temperature (LSAT) series (e.g. Dieng et al. 2017, Chan et al. 2023, Ishii et al. 2025). Certain features of the global historical record where land and sea temperature components do not agree are being attributed with increasing confidence to residual biases in SST datasets (e.g. Chan et al. 2023, Sippel et al. 2024). This includes the Second World War (WW2) warm anomaly: anomalous warmth found in the SST record between 1941 and 1945 due to rapid changes in measurement practises, that persists in published SST datasets despite existing bias correction procedures (e.g. Chan and Huybers 2021). This feature is present in both SST and marine air temperature datasets (Kent et al. 2013, Cornes et al. 2020), but not in the land near-surface air temperature record.

To address these residual biases, some recent SST datasets have incorporated methods of aligning the SST record more closely with near-surface land air temperatures. COBE-SST3 (Ishii et al. 2025) calculates SST bias corrections using a version of HadNMAT2 prior to 2010 which, before use, is adjusted according to its global differences from the CRUTEM5 near-surface land air temperature dataset (Osborn et al. 2021), between 1890 and 1950 when these differences are largest. The DCENT dataset (Chan et al. 2024b) uses a coupled energy-balance model to cross-calibrate its land (DCLSAT) and sea (DCSST) components (Chan et al. 2024a).

HadSST.4.1.1.0 follows a physically-based approach to bias correction as described in Kennedy et al. (2019). Uncorrected SST observations are averaged onto a 5° latitude-longitude grid, along with the fraction of different measurement types (bucket, ERI, buoy or "unknown") contributing to the average in each grid cell. An ensemble of bias estimates for each measurement type is then combined with an ensemble of measurement type fractions for each grid box (accounting for both unknown and potentially mis-reported metadata) to produce an ensemble of overall bias corrections. This approach makes use of nighttime marine air temperature (NMAT) and near-surface ocean temperature measurements for bias estimation, but remains independent of the land temperature record.

In this article we describe and evaluate an update of HadSST that addresses the WW2 warm anomaly. Section 2 quantifies the WW2 warm anomaly in HadSST4, identifies contributing factors and proposes a solution. Section 3 documents further changes that have been implemented in HadSST4 since the most recent publication (HadSST.4.0.0.0, Kennedy et al. 2019). Section 4 then compares the new HadSST.4.2.0.0 release with the previous version (4.1.1.0), and discusses similarities and differences from other SST datasets including COBE-SST3 (Ishii et al. 2025), DCSST (Chan et al. 2024a) and ERSSTv6 (Huang et al. 2025a and 2025b).

## 2. Anomalous warmth in HadSST4, 1941-1945

Uncorrected SST data show a significant warm bias during the WW2 period. This is due to the rapid introduction of engine room intake (ERI) as the primary source of SST measurements during wartime. Since ERI measurements are typically biased warm, and buckets cool, this leads to a significant discontinuity in observed temperatures at the start of the Second World War.

Chan and Huybers (2021) propose a quantitative definition of the WW2 warm anomaly, the procedure for calculating which is:

1. Calculate global annual average temperature anomalies for the Second World War (1941-1945 inclusive) and the five year periods immediately before (1936-1940) and after (1946-1950).
2. Calculate the midpoint between the "before" and "after" period averages. The WW2 warm anomaly is the difference between the 1941-1945 average and this midpoint.

Using this definition, the authors quantify WW2 warm anomalies in various datasets including ERSST versions 4 and 5, and HadSST versions 2-4. The HadSST.4.0.0.0 ensemble is found to have an average warm anomaly of 0.19°C, with the 200 individual members ranging from -0.09°C to 0.45°C. By generating and comparing SST reconstructions from different subsets of measurements, both bias-corrected and uncorrected, Chan and Huybers (2021) estimate the "true" WW2 warm anomaly to have been in the range -0.01°C to 0.18°C.

## Causes of WW2 warm anomaly in HadSST4

The current version of the HadSST dataset (4.1.1.0) has a median WW2 warm anomaly of 0.21°C with a range of -0.11°C to 0.48°C. This indicates under-correction for warm bias during the WW2 period. In the HadSST4 framework (see Kennedy et al. 2019) this could occur through one of two mechanisms: underestimation of the ERI bias, or underestimation of the fraction of ERI measurements during WW2.

HadSST4 introduced a method of estimating grid-box ERI biases with respect to near-surface subsurface (HadIOD; Atkinson et al. 2014) and buoy measurements (Kennedy et al. 2019). However, this method is applied only from 1950 onwards, due to the paucity of data in the early period. Prior to 1950, a single value of ERI bias is assumed to apply globally. An ensemble of bias values is sampled from a uniform distribution between 0.1 and 0.5°C prior to 1950 (Kennedy et al. 2011), and from 0.0 to 0.5°C during WW2 (1941-1945 inclusive; Kennedy et al. 2019).

Figure 1 shows the relationship in the HadSST.4.1.1.0 ensemble between WW2 warm anomaly and the global ERI bias assumed during WW2. The relationship is almost linear, with a correlation of -0.93. HadSST.4.1.1.0 members with a WW2 warm anomaly falling within the estimated "true" range (-0.01°C to 0.18°C, Chan and Huybers 2021) are those where the global ERI bias is estimated to be in the range 0.2-0.5°C, and every member using an ERI bias of less than 0.2°C has anomalous warmth during WW2. While Chan and Huybers (2021) infer that the fraction of unknown measurements assumed to be from buckets is responsible for the re-emergence of a WW2 warm anomaly in HadSST4, no correlation was found between these two parameters in HadSST.4.1.1.0; instead there is a clear influence from the ERI bias value.

In quantifying the "true" WW2 warm anomaly which strongly correlates with ERI bias value, Chan and Huybers (2021) provide what is effectively a new constraint on the ERI bias during WW2. The correlation is not perfect, but this constraint would substantially increase the proportion of ensemble members with WW2 warmth within the expected range. In light of this, we review the evidence for the pre-1950 ERI bias ranges assumed in HadSST.4.1.1.0 and consider whether it is justifiable to update them.

## Revisiting WW2 ERI bias estimates

The origins of the pre-1950 ERI bias estimates used in HadSST4 are explained in Kennedy et al. (2011). The authors review existing literature quantifying ERI biases at different points in time, summarising their findings in a table that states the size and period of the study. Figure 2 shows these sources plotted by the dates for which ERI bias was estimated, and colour-coded by the size of the study. Overplotted on these are the current and proposed pre-1950 ERI bias correction ranges, and globally averaged timeseries of analysed ERI biases from HadSST.4.1.1.0.

In justifying their estimates of early ERI bias Kennedy et al. (2011) state that: "The typical bias is around 0.2°C and most estimates fall between 0.1 and 0.3°C. The larger biases are typically from small samples of ships…" However, Figure 2 shows that the larger studies providing more reliable – and typically smaller – ERI bias estimates all date from the more recent period, with none of the largest studies (plotted in green) dating from prior to 1960. There is also clear evidence from bias analyses (calculated relative to subsurface and buoy data as described in Kennedy et al. (2019)) that ERI biases are not static, but have decreased over time since the 1960s. In fact, there are very few estimates available of ERI biases prior to 1950, all of which are from very small studies and with substantial variation between estimates. These studies also tend to quantify biases with respect to buckets (either "tin" or simply "bucket"), which have biases of their own.

The uncertainties on ERI bias values are typically not quoted in the references of Kennedy et al. (2011). However, James and Fox (1972) found a very large spread in measurements: their mean bias was 0.3°C, but with a 1σ range of around ±0.9°C based on factors such as ERI thermometer positioning, weather conditions (wind speed, precipitation and air-sea temperature difference), ship speed, and the type of bucket used as the reference. Kent and Kaplan (2006b) also found random error variances of 2.7-3.0(°C)$^2$ on their ERI bias estimates between 1975 and 1994. Both of these data points (although not their uncertainties) are included in Figure 2. Assuming that the uncertainties on single ship estimates are similarly large, we submit that the literature reviewed by Kennedy et al. (2011) actually provides very little constraint on likely ERI biases prior to the 1960s, when larger studies start to become available.

We propose that a pre-1950 ERI bias range of 0.2-0.5°C, as inferred in this paper from the expected size of the WW2 warm anomaly (Chan and Huybers 2021), is no less consistent with the existing literature than the ranges currently used in HadSST.4.1.1.0. We therefore introduce and test this new range in HadSST.4.2.0.0.

# 3. Additional developments in HadSST.4.2

## Technical reimplementation (HadSST.4.1.0.0)

The HadSST.4.1.0.0 dataset was released in February 2025 and represented a full technical re-implementation of the HadSST4 system. The original IDL system was re-coded in modular Python, and manual processing was replaced by automated Cylc 8 workflows (Oliver et al. 2019), substantially reducing the risk of human error during the monthly update process.

While the science of HadSST.4.1.0.0 remains as documented by Kennedy et al. (2019), the process of re-implementation uncovered some bugs in the original IDL which were fixed as part of the release. The impact on global and hemispheric SST anomalies, ensemble median and uncertainties was negligible. However, there was a change to an intermediate result: the inferred dates for the transition between uninsulated canvas and insulated rubber buckets in the mid-to-late 20$^{th}$ century.

Kennedy et al. (2019) describe a method of inferring plausible start and end dates for the canvas-to-rubber bucket transition using the analysed biases of different subsets of the input observations (bucket, ERI, and "all ship" measurements). The "accepted" start and end dates define a parameter space from which 200 realisations are sampled for the ensemble. The transition dates determined in HadSST.4.0.1.0 are shown in the left-hand panel of Figure 3, which was published as figure 7c in Kennedy et al. (2019). The implication is that insulated rubber buckets could have been starting to be introduced as early as the 1930s, with possible start dates prior to the Second World War (up to and including 1940) representing 29% of the sampling space. The changes introduced in HadSST.4.1.0.0, however, lead to a more restricted subset of start and end dates and a later transition (right-hand panel of Figure 3), with only 1.1% of possible transitions starting prior to 1941.

Of existing global temperature and SST datasets, very few make an explicit distinction between bucket types or attempt to define their proportions for different periods in history. The most thorough discussion of the timing of the canvas-to-rubber bucket transition is found in Kennedy et al. (2011), which considers the guidelines for SST measurements provided by several meteorological agencies between the 1920s and 1980s. The authors conclude that rubber buckets were most likely starting to be introduced in the mid-1950s, with the transition away from canvas buckets completed between 1970 and 1980.

A move towards rubber buckets in the 1950s rather than the 1930s is consistent with the wider literature. Kent et al. (2010), Kent and Taylor (2006b) and Folland and Parker (1995) all discuss SST biases in terms of a mixture of canvas and wooden buckets prior to the Second World War. Hirahara et al. (2014) estimate insulated bucket fractions for COBE-SST2 for dates between 1942 and 2010, implicitly assuming that insulated (rubber) buckets were not widespread before that time. Both Hirahara et al. (2014) and Carella et al. (2018) also find evidence that canvas bucket biases dominate overall SST bias estimates until at least the 1960s, supporting a later transition than suggested by Kennedy et al. (2019). The changes made in HadSST.4.1.0.0 therefore bring this result into closer agreement with the wider literature.

## Truncation biases in the Kobe collection

HadSST4 uses in-situ observations from the International Comprehensive Ocean-Atmosphere Data Set (ICOADS; Freeman et al. 2017) prior to 2021 (Kennedy et al. 2019). A known truncation bias exists in data from the Japanese Kobe Collection, where SST measurements in ICOADS deck 118 were truncated to the nearest whole degree on reporting (Chan et al. 2019; Ishii et al. 2025). Chan and Huybers (2019) find that the Kobe Collection as a whole (comprising ICOADS decks 118, 119 and 762) is cold-biased by 0.4°C with respect to measurements from other decks.

In the COBE-SST3 dataset (Ishii et al. 2025) all deck 118 measurements are adjusted upwards by 0.5°C to correct for truncation bias. This assumes that, on average, the "true" value of truncated measurements will have been uniformly sampled from the distribution 0-1°C above the reported value. In practise this distribution does not extend all the way to 1, but depends on how precisely the thermometer can be read. The reference manual for deck 118, retrieved from the RECLAIM project website (https://icoads.noaa.gov/reclaim/pdf/dck118.pdf, accessed 20/06/2025; Wilkinson et al. 2011) states that for SST: "1/10°C values are dropped and punched in whole °C". We therefore assume that the thermometer would typically be read to around the nearest 1/10°C, giving a central truncation bias value of 0.45°C, consistent with Chan et al. (2019). We add this value to the deck 118 measurements.

In addition to that in deck 118, a lesser-known bias appears to exist in measurements from deck 119. A Met Office Discussion Note (Parker 1989) records the following:

> *The description of Tape Deck 1119 (Japanese Ship Observations 1953-1961) states that sea temperature was computed from air temperature (in whole °C) and air minus sea temperature (in halves °C), and that, during this computation, decoded values of the latter were not rounded,* *i.e. they were truncated. This would make SST too low by 0.25°C on average where the sea is warmer than the air.*

The reference manual for deck 119 (https://icoads.noaa.gov/reclaim/pdf/dck119.pdf, accessed 20/06/2025) is consistent with Parker (1989) in that it confirms the rounding of air temperature and air-sea temperature difference in the original measurements, but it does not contain details of the SST computation perfomed when the records were digitised.

In a more recent paper discussing Kobe Collection impacts on early 20[th] century temperature trends, Chan et al. (2019) mention that in addition to deck 118, deck 119 is "also truncated" and cold-biased. However, this was not intended to imply the same mechanism of truncation as for deck 118 (Duo Chan, pers. comm., 1 September 2025). The impact of the Parker (1989) mechanism in a context where the sea is frequently warmer than the air would also be consistent with a small cold offset in deck 119 measurements, as observed by Chan et al. (2019).

We examined the ICOADS.3.0.0 data (Freeman et al. 2017) used in HadSST4 (1850-2014; Kennedy et al. 2019) and determined that the SST data from decks 118 and 119, and only these decks, are integer-only. This provides evidence for truncation effects in deck 119 as well as deck 118, and is consistent with the observations of both Parker (1989) and Chan et al. (2019). On this basis, we adjust deck 119 SSTs upwards by 0.25°C where SST is higher than the reported air temperature, and downwards by 0.25°C where SST is lower. Where the two are equal, no correction is required.

# 4. Results and discussion

## Comparison with HadSST.4.1.1.0

Figure 4 shows the annual global mean anomaly of the HadSST.4.2.0.0 median compared to the previous version (HadSST.4.1.1.0), with the 95% total uncertainty range (encompassing uncertainties from correlated and uncorrelated measurement errors, sampling uncertainty, coverage uncertainty, and bias correction uncertainty as represented by the 200-member ensemble). The most significant changes are a reduction of order 0.1°C in global median SST anomaly during the WW2 period, from 1941 to 1945 (inclusive), and an increase of similar magnitude between 1930 and 1941. The total uncertainty during WW2 is reduced in HadSST.4.2.0.0, due to the narrower range of ERI bias corrections (smaller bias correction uncertainty represented by reduced ensemble spread), making the uncertainty during this period more comparable to that of neighbouring periods.

Table 1 shows the number of members with an anomalously warm or cool WW2 period in HadSST.4.2.0.0 compared with the previous version. The combined impact of the truncation bias correction to deck 118 (increasing global temperatures 1930-1941) and increasing the estimated ERI warm bias prior to 1950 (reducing global temperatures 1941-1945) results in only 31 realisations of the 200-member ensemble retaining an anomalous warmth during WW2, compared to over half (106) of the members in HadSST.4.1.1.0. 80% (160 members) of the HadSST.4.2.0.0 ensemble now lie within the estimated "true" range for this period as defined by Chan and Huybers (2021). This is also illustrated in the narrower distribution of WW2 anomaly values in Figure 5, centred within the "true" range. The fact that 20% of members still lie outside of this range (15% warm, 5% cool) is consistent with the correlations between the WW2 warm anomaly and assumed ERI bias found in HadSST.4.1.1.0 (Figure 1), and may indicate further factors contributing to the residual anomaly.

Outside the WW2 period, there is a significant increase in global mean SST anomaly between 1930 and 1941. This is attributable to the truncation bias correction of deck 118, which contributes a large number of global observations during this period (Figure 6) and a significant proportion of observations from the North Pacific and Indian Oceans (Figure 7). By contrast, contributions from deck 119 in the late 1950s (Figure 6) are limited in both proportion and extent (Figure 8), and truncation corrections do not noticeably impact the global mean.

The significance of the contribution of deck 118 can be seen in the impact on decadal averages in the 1930s (Figure 9). HadSST.4.2.0.0 is warmer than HadSST.4.1.1.0 in this decade across widespread regions of the globe. The largest impact, as expected, is seen in the North Pacific, but increases of 0.1°C or more are visible in the South-West Pacific, Indian Ocean and parts of the Southern Ocean. The overall change in global SST anomaly during this period is consistently of order 0.1°C.

Figure 10 shows the change to decadal mean SST anomalies in the 1940s. Grid box reductions of up to 0.1°C are widespread, consistent with the increase in ERI bias correction over the first half of this decade. However, there is significant regional variation, with localised increases to temperatures particularly in the far North and North-West Pacific due to the influence of deck 118.

Selected regional differences in the impact of the HadSST.4.2.0.0 developments are shown in Figure 11. Increased SST anomalies due to the adjustment of deck 118 prior to 1941 are concentrated in the Northern Hemisphere, with smaller contributions to the Southern Hemisphere from Indian and Southern Ocean effects. All regions show a cooling from 1941 through 1945 due to the increased ERI bias correction, which leads to a flatter temperature profile globally and in the Northern Hemisphere regions (North Atlantic and North Pacific). The combined impacts in the North Pacific in particular lead to a temperature profile that is distinctly different in HadSST.4.2.0.0 from the previous version, with substantially less variation in temperature anomalies over the 1930s and 40s. By contrast, impacts in the North Atlantic are due only to the changes in WW2 ERI bias, as the contribution of Kobe Collection observations in this region is negligible (Figure 7 and Figure 8).

A new pattern of temperature trends emerges in the Southern Hemisphere timeseries in HadSST.4.2.0.0, with a slight but steady cooling between 1940 and 1946. This contrasts with the pattern in HadSST.4.1.1.0 of steady warming to 1944, followed by a more abrupt cooling at the end of the WW2 period. A new temperature peak emerges in 1940 in the Southern Hemisphere and Indian Oceans from the large contribution of deck 118 measurements at that time.

HadSST.4.2.0.0 gives a qualitatively different story as to the evolution of SSTs between the early 1930s and 1950, compared to HadSST.4.1.1.0. To place these in context, Figure 12 shows Niño 3.4 indices computed for both versions of HadSST (which are almost identical), plotted with the Southern Oscillation Index (SOI; Ropelewski and Jones 1987) for the period of interest. Strong negative peaks in SOI corresponding with a positive Niño index are generally associated with warmer global temperatures, and vice versa. Both 1940 and 1941 were El Niño years with a strongly negative SOI, consistent with the new peak in SST anomalies in 1940 (Figure 11). Both indices support local SST anomaly minima in 1938, 1943 and 1950 that are all found in HadSST.4.2.0.0, but HadSST.4.1.1.0 lacks the local minimum in 1943. There is poor agreement between the SOI and Niño 3.4 indices in the second half of the WW2 period, when there was extremely poor spatial coverage of SST observations in the Niño 3.4 region (Figure 12).

Since the WW2 warm anomaly is typically associated with marine datasets only, Figure 13 compares the two HadSST versions with selected land temperature datasets between 1920 and 1970. DCLSAT is the land component of the DCENT dataset (comprising DCLSAT and DCSST; Chan et al. 2024a and 2024b), while CRUTEM5 is an established gridded land near-surface air temperature dataset (Osborn et al. 2021). GloSATLAT is the land air temperature component of the GloSAT reference analysis v1.0.0.0 (Morice et al. 2025).

During the WW2 period, HadSST.4.2.0.0 shows better qualitative agreement with land temperature datasets than HadSST.4.1.1.0 (Figure 13). Figure 12Specifically, the land datasets and HadSST.4.2.0.0 all show a dip in temperatures in 1942 and 1943 that is not present in HadSST.4.1.1.0. Additionally, the three land temperature series are qualitatively consistent with the changed "story" of temperature evolution in HadSST.4.2.0.0: a slow increase in temperatures through the 1930s, followed by a gradual cooling from the late 1930s to 1950. This contrasts with the much later peak and sudden temperature drop after 1945 in HadSST.4.1.1.0.

## Comparison with other SST datasets

To place these changes in a wider context, we compare global and regional averages from the new version of HadSST with other recently updated global SST datasets: DCSST (Chan et al. 2024a), COBE-SST3 (Ishii et al. 2025) and ERSSTv6 (Huang et al. 2025a and 2025b). To facilitate comparison, datasets were first regridded onto the HadSST 5° monthly grid using an area-weighted averaging scheme. Each 5° grid cell was baselined to represent an anomaly with respect to its own 1961-1990 climatology. Datasets were then reduced to areas of shared spatial coverage before calculating regional timeseries.

Figure 14 shows annual global mean anomalies for the DCSST ensemble mean, COBE-SST3, ERSSTv6, HadSST.4.1.1.0 and HadSST.4.2.0.0 medians from 1850 to 2024. The datasets show close agreement after 1970, but diverge in earlier periods, with substantial differences prior to 1920. While HadSST, COBE and ERSST are all warmer in the late 19[th] century, with a cooler period between around 1890 and 1920, DCSST shows a more gradual and consistent increasing trend through the 19[th] and early 20[th] centuries.

The differences in dataset evolution are due to structural differences in dataset creation that affect the early period. There is substantial uncertainty in the biases affecting bucket SST measurements prior to WW2 (e.g. Folland and Parker 1995), with increasing evidence for a residual cold bias in SSTs in the early 20[th] century (Chan et al. 2023, Sippel et al. 2024, Chan et al. 2025). Of the datasets with a cool period 1890-1920, COBE and ERSST use estimated bucket biases to correct individual observations for absolute offsets, while HadSST adjusts gridded SST anomalies based on the estimated contributions of different measurement methods. By contrast, DCSST first adjusts observations in a relative fashion to bring nearby SST estimates into closer agreement (Chan et al. 2021), and then uses a coupled energy-balance model to derive coastal calibration factors (Chan et al. 2023), drawing sea surface temperature evolution into closer agreement with the land component of the full DCENT dataset. The qualitatively different storylines produced by these different approaches reveals large structural uncertainties in global pre-industrial SST estimates.

Table 2 quantifies WW2 warm anomalies for each of the five global SST datasets. Both DCSST and COBE-SST3 have anomalies close to zero. In both cases this is achieved by calibrating the SST timeseries (indirectly) against land temperatures, to compensate for features likely attributable to residual SST biases. COBE-SST3 applies a two-step method to estimate bias corrections for SST measurements using HadNMAT2 (Kent et al. 2013), which is itself first calibrated against CRUTEM5, as described in Ishii et al. (2025). In DCENT, the land calibration is applied via a coastal coupled energy-balance model (Chan et al. 2023). By contrast, ERSSTv6 retains a WW2 warm anomaly of 0.20°C – reduced from 0.29°C in ERSSTv5 (Chan and Huybers 2021), but close to the HadSST.4.1.1.0 value of 0.21°C. ERSSTv6 performs bias adjustments based on HadNMAT2 prior to 2010, but without the adjustment towards CRUTEM; and night marine air temperature (NMAT) datasets are also known to show anomalous warmth during the WW2 period (e.g. Kent et al. 2013; Cornes et al. 2020). In HadSST.4.2.0.0 we halve the size of the WW2 warm anomaly by improving our estimates of ERI bias prior to 1950, while preserving the independence of SSTs from land temperature series. The HadSST.4.2.0.0 median WW2 warm anomaly of 0.11°C is close to the centre of the "true" range estimated by Chan and Huybers (2021).

Figure 15 shows annual global and regional anomaly timeseries for the five SST datasets (DCSST, COBE-SST3, ERSSTv5, HadSST.4.1.1.0 and HadSST.4.2.0.0) between 1920 and 1970. At this resolution, the anomalous warmth from 1941 to 1945 in ERSSTv6 is clearly visible in comparison with other datasets. The main difference between datasets with and without overall anomalous warmth is the size of the temperature peak in the earlier half of WW2. All datasets have a warm peak towards the end of WW2 (1944 and 1945), which is least pronounced (but still present) in DCSST. It is not clear from the Niño 3.4 or Southern Oscillation Indices whether or not a global temperature peak would be expected here: the 1944 peak in SOI is weak and the Niño region at that time very poorly sampled (Figure 12).

Regionally, HadSST.4.2.0.0 and COBE-SST3 are warmer than other datasets for the globe, Northern Hemisphere and North Pacific from the mid-1930s until WW2 (Figure 15). This is particularly noticeable in the North Pacific 1933-1940, where deck 118 is dominant. ERSSTv6 is coolest in these three regions, and DCSST lies between them. This is likely due to the fact that both COBE-SST3 and HadSST.4.2.0.0 apply a direct truncation bias correction to deck 118 measurements, while DCSST derives a relative adjustment via groupwise bias offsets and coastal calibration. We can infer that the effective adjustment in DCSST is slightly smaller than those applied to the individual deck (0.5°C for COBE-SST3 (Ishii et al. 2025), and 0.45°C for HadSST.4.2.0.0).

There are regions where DCSST in particular tells a different story from other SST datasets. In the North Pacific, DCSST has a peak in 1930 followed by lower temperatures than HadSST.4.2.0.0 and COBE-SST3 in the late 1930s and during WW2. This gives a picture of SSTs that are relatively consistent through the 1930s and 40s. By contrast, HadSST.4.2.0.0 and COBE-SST3 peak later in the 1930s and remain higher overall during this period. In the North Atlantic, DCSST is the warmest dataset in the 1920s and coolest in the 1940s. In the Southern Hemisphere, DCSST evolves similarly to HadSST.4.2.0.0 and COBE-SST3 everywhere except for the WW2 period. These differences reflect the very different way in which DCSST is constructed compared to the other datasets.

Figure 16 shows the differences of each of the other SST anomaly datasets (DCSST, COBE-SST3 and ERSSTv6) from HadSST.4.2.0.0. In general, HadSST is most similar to COBE-SST3, reflecting the fact that the COBE and HadSST bias corrections are conceptually very similar. Before 1939, bucket SST observations in COBE are corrected according to the Folland and Parker (1995) scheme, and a 0.5°C adjustment for truncation bias is applied to deck 118. For later observations, spatially-varying bucket and ERI biases are estimated with respect to Argo and buoy data. Therefore the agreement is unsurprising, with the main difference being the 1940 peak in HadSST.4.2.0.0 that is absent from COBE-SST3. However, COBE-SST3 also adjusts bias-corrected SSTs towards CRUTEM5 (indirectly via HadNMAT2) in the period between 1890 and 1950 (Ishii et al. 2025), which includes the WW2 period. By contrast, the HadSST.4.2.0.0 approach of updating ERI bias estimates remains physically-based and independent of land temperature datasets, and effectively addresses anomalous warmth in WW2.

Globally, all three SST datasets are within total uncertainty from the HadSST.4.2.0.0 median for the majority of the period shown in Figure 16, with the exception of ERSSTv6 from the mid-1920s until 1940. The differences between HadSST.4.2.0.0 and ERSSTv6 are much smaller and less variable over time in the North Atlantic, supporting an inference that the pre-WW2 differences elsewhere are largely due to the treatment of deck 118. DCSST also shows local differences from HadSST.4.2.0.0 including in the North Pacific prior to WW2, that lie outside of the expected uncertainty range. These are likely due to the very different bias correction approach in DCSST, and again highlight large structural uncertainties in SSTs at the regional level.

Figure 17 shows the difference between the five SST datasets above and CRUTEM5. This emphasises the close agreement between HadSST.4.2.0.0 and COBE-SST3 over the period of interest, with both showing similar differences from CRUTEM5, despite the fact that COBE-SST3 uses CRUTEM5 in its construction and HadSST.4.2.0.0 does not. DCSST tracks differently from the other datasets, matching the land series on average more closely in the late 1930s and 1940s. ERSSTv6 shows the largest deviation and the least temporal consistency, being noticably cooler than CRUTEM5 in the 1920s and 1930s, but having the largest warm peaks in the WW2 period. ERSSTv6 is also the coolest of the SST datasets in the late 1940s, exacerbating its WW2 warm anomaly.

# 5. Conclusions

This paper presents an update to the HadSST dataset incorporating technical improvements, corrections for truncation biases in the Japanese Kobe Collection (1933-1961), and removal of the WW2 warm anomaly. By using the Chan and Huybers (2021) definition of the WW2 warm anomaly and their estimate of its likely "true" value, we propose new constraints on ERI biases prior to 1950 that are consistent with the previous literature. Using these new ERI bias estimates reduces the median WW2 warm anomaly from 0.21°C in HadSST.4.1.1.0 to 0.11°C in HadSST.4.2.0.0, with 80% (160 members) of the new ensemble lying within the expected anomaly range (-0.01 to 0.18°C, Chan and Huybers 2021).

Maintaining a physically-based approach to resolving biases in early SST data is unusual in the context of recent literature. Both COBE-SST3 (Ishii et al. 2025) and DCENT (Chan et al. 2024a) have implemented statistical approaches that effectively adjust bias-corrected SSTs to more closely match land near-surface air temperature records. While this reduces the appearance of residual biases in the early 20$^{th}$ century (Chan and Huybers 2021, Chan and Huybers 2023, Sippel et al. 2024, Chan et al. 2024a, Ishii et al. 2025), it introduces a dependence between land and sea surface temperature records, and may ultimately reduce structural diversity leading to an underrepresentation of uncertainties across the range of SST and global near-surface temperature datasets. In this paper we have shown that it is possible to correct at least the WW2 warm anomaly within a bias model that remains independent of land temperature records. This complements state-of-the-art statistical approaches and allows this important element of structural diversity to be retained among the most up-to-date SST datasets.

## Acknowledgements

The authors are grateful to John Kennedy for his time and expertise in explaining the workings of HadSST.4.0.0.0. We would also like to thank Elizabeth Kent for sharing her knowledge of ICOADS, including details of the original tape decks to support our investigation into the Kobe Collection truncation biases. Caroline Sandford and Nick Rayner were supported by the Met Office Hadley Centre Climate Programme funded by DSIT.

## Author contributions

**Caroline Sandford:** Software; writing – original draft; writing – review and editing; visualization; data curation.

**Nick Rayner:** writing – review and editing; supervision; resources.

## Data availability statement

The HadSST.4.1.1.0 and 4.2.0.0 datasets are publicly available via the Met Office Hadley Server: https://www.metoffice.gov.uk/hadobs/hadsst4/.

# Tables

*Table 1: WW2 warm anomaly for HadSST.4.1.1.0 and HadSST.4.2.0.0, according to the definition of Chan and Huybers (2021). Number of members with anomalies above, below and within the expected range ("no anomaly"), anomaly of the ensemble median, and range of anomalies across the 200 HadSST ensemble members.*

|  | **HadSST.4.1.1.0** | **HadSST.4.2.0.0** |
|---|---|---|
| **Members with warm WW2** | 106 | 31 |
| **Members with cool WW2** | 17 | 9 |
| **Members with no WW2 anomaly** | 77 | 160 |
| **WW2 anomaly of ensemble median (°C)** | 0.21 | 0.11 |
| **Median (range) of WW2 anomaly (°C)** | 0.21 (-0.11 – 0.48) | 0.11 (-0.10 – 0.26) |

*Table 2: WW2 warm anomalies for a range of historical sea-surface temperature (SST) datasets according to the definition of Chan and Huybers (2021).*

| **Marine temperature series** | **WW2 warm anomaly (°C)** |
|---|---|
| **HadSST.4.1.1.0 ensemble median** | 0.21 |
| **HadSST.4.2.0.0 ensemble median** | 0.11 |
| **DCSST ensemble mean** | 0.06 |
| **COBE-SST3** | 0.06 |
| **ERSSTv6** | 0.20 |

# Figures

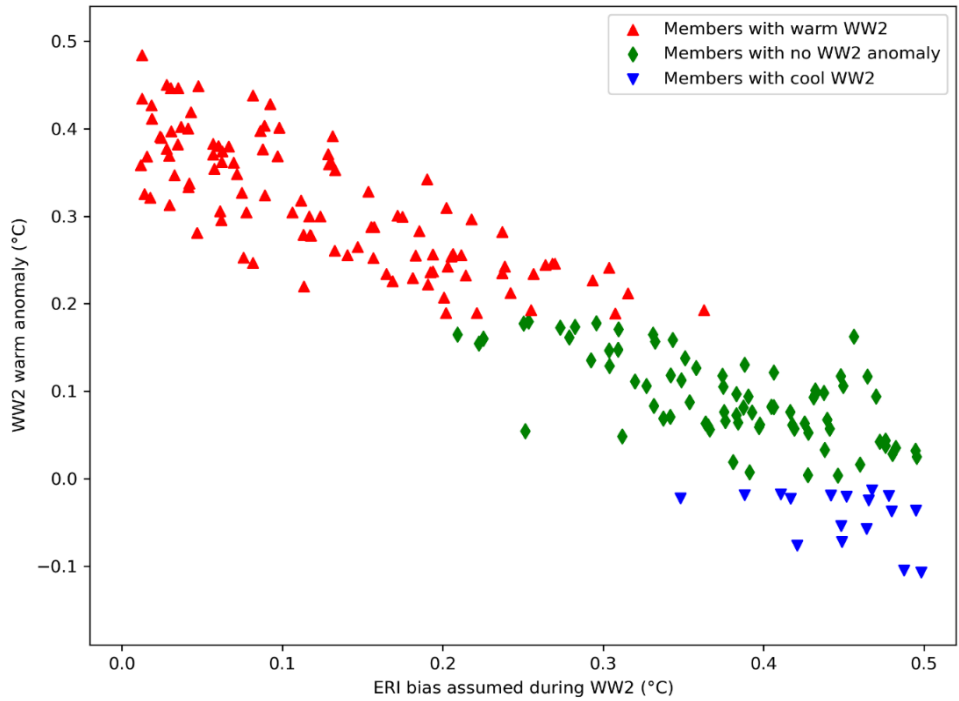

*Figure 1: WW2 warm anomaly for each member of the HadSST.4.1.1.0 ensemble, plotted against the ERI bias assumed during WW2. Members are classified as having a warm, cool, or no WW2 anomaly based on the expected "true" anomaly range (-0.01 to 0.18°C) of Chan and Huybers (2021).*

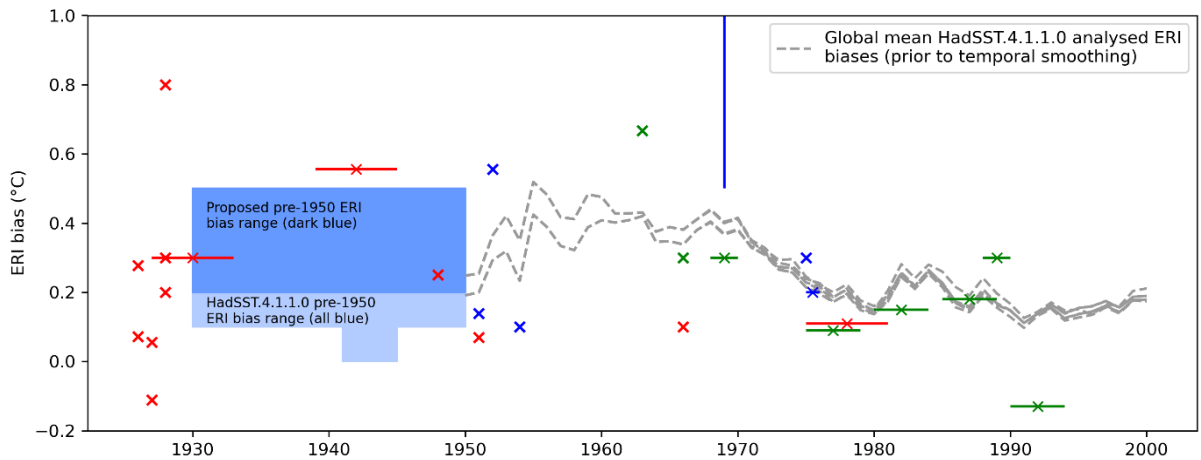

*Figure 2: Graphical version of Kennedy et al. (2011)'s table 3, showing ERI bias estimates and the period for which they were calculated. Studies are colour-coded by the quantity of observations analysed, with red crosses indicating studies based on data from a single ship, and green crosses for studies that used at least several thousand observations. Studies of intermediate quality (more than one, but still few ships) are shown in blue. This includes one study that reported an ERI bias range of 0.5-2.3°C (errorbar overshoots the y-axis) against an unknown reference.*

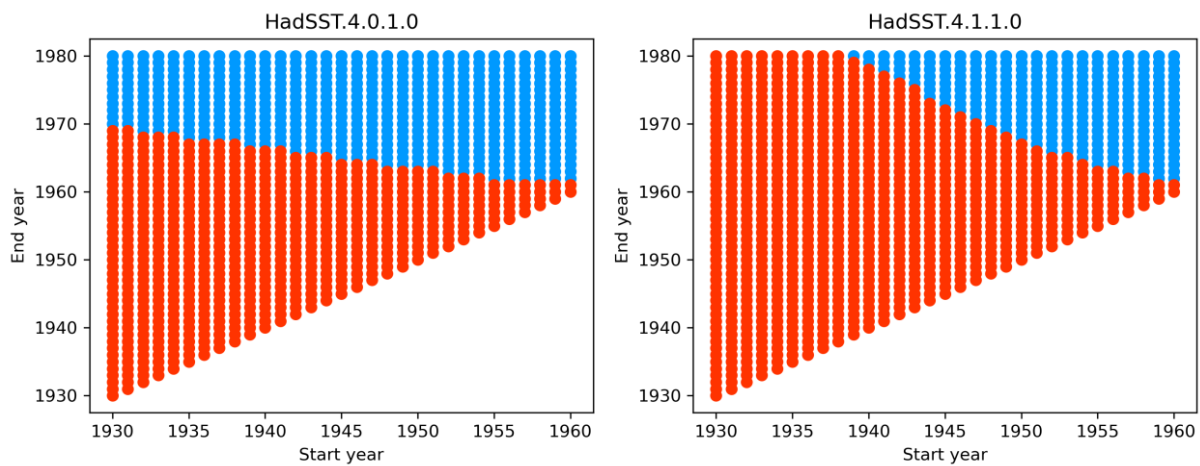

*Figure 3: Accepted (blue) and rejected (red) start and end dates for the canvas to rubber bucket transition in HadSST4. Left: for HadSST.4.0, published as figure 7 in Kennedy et al. (2019). Right: after fixes applied in HadSST.4.1.*

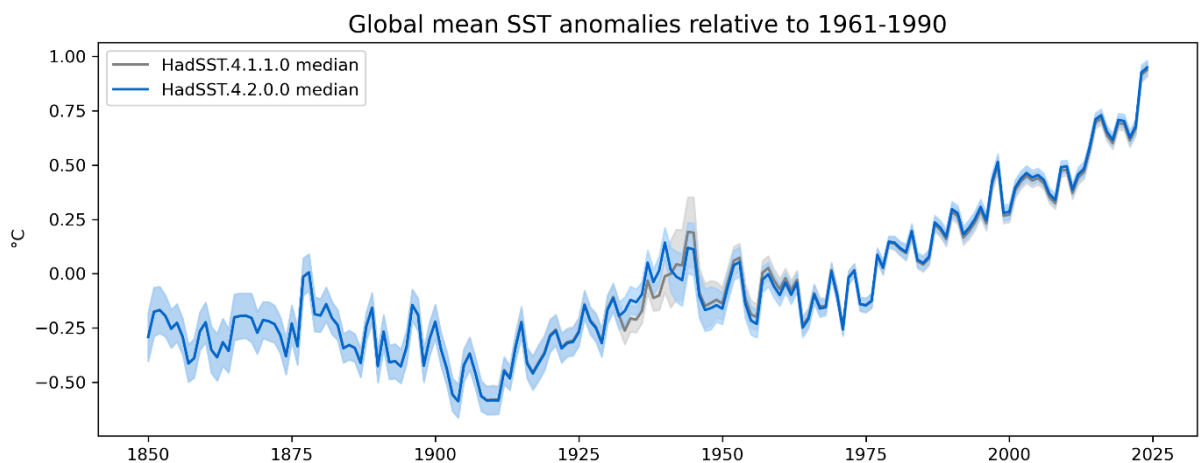

*Figure 4: Annual global mean sea surface temperature (SST) anomaly medians for HadSST.4.2.0.0 and HadSST.4.1.1.0, with shaded total uncertainty bounds incorporating correlated and uncorrelated measurement uncertainties, sampling, coverage and bias correction uncertainties. The annual anomaly is the mean of the 12 months' anomalies from each year.*

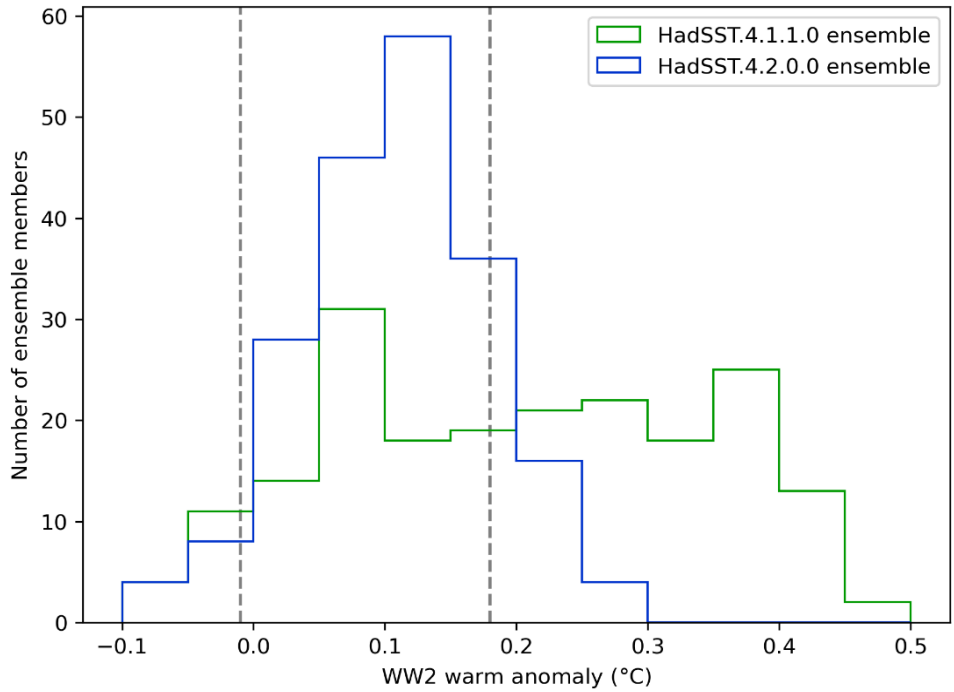

Figure 5: Distribution of WW2 anomaly values across the 200 members of the HadSST.4.2.0.0 and HadSST.4.1.1.0 ensembles. Dashed vertical lines at -0.01°C and 0.18°C show the uncertainty range on the estimated "true" anomaly according to Chan and Huybers (2021).

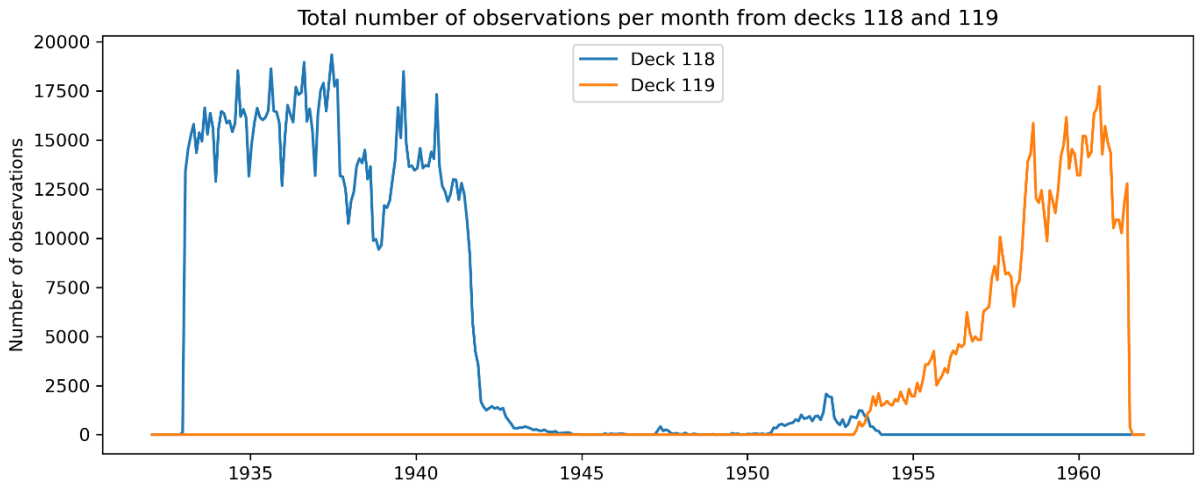

Figure 6: Total number of observations per month from decks 118 and 119 that are used to inform the HadSST gridded anomalies.

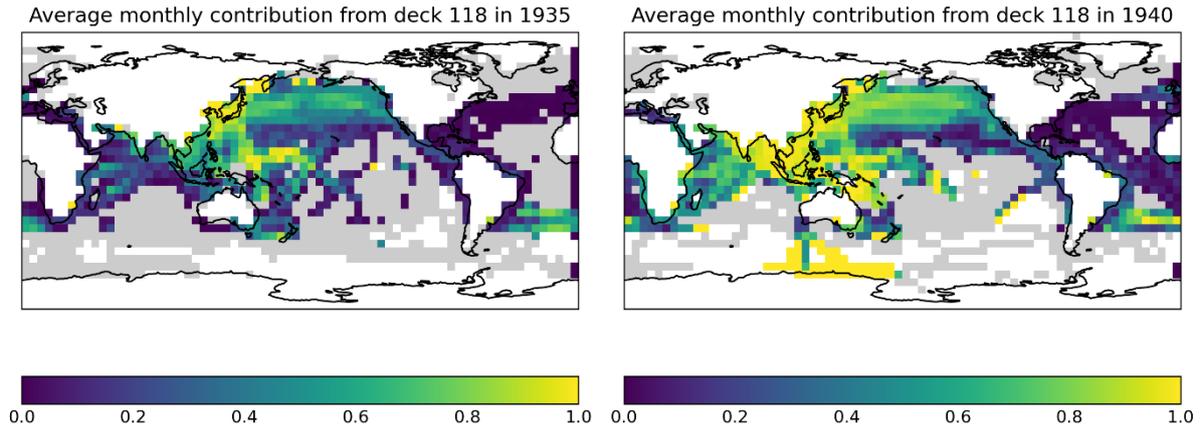

*Figure 7: Proportional contribution of deck 118 measurements to each 5° grid cell, averaged over the 12 months in 1935 (left) and 1940 (right).*

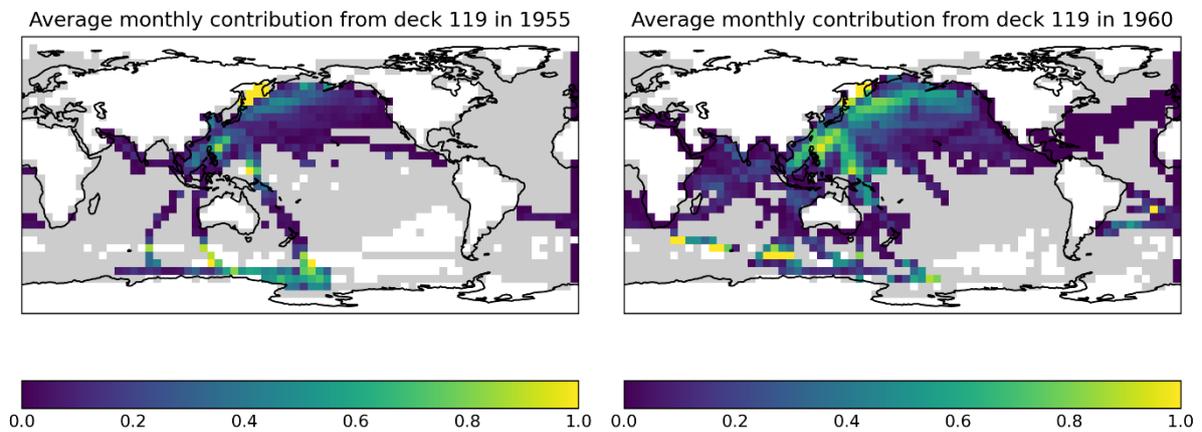

*Figure 8: Proportional contribution of deck 119 measurements to each 5° grid cell, averaged over the 12 months in 1955 (left) and 1960 (right).*

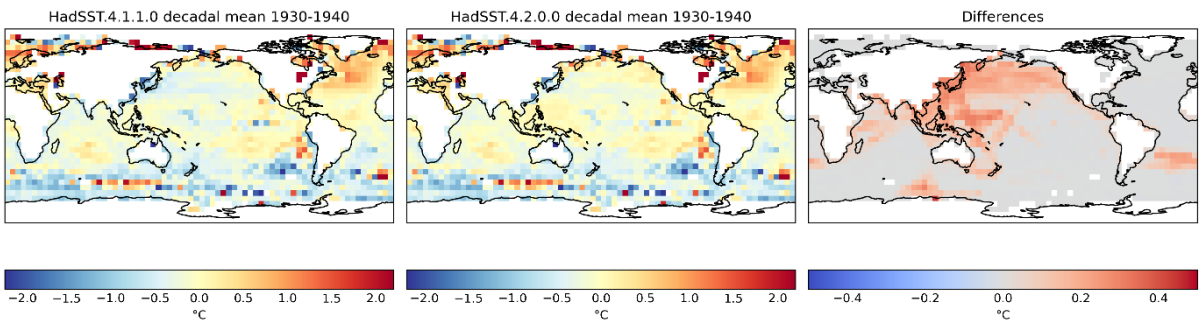

*Figure 9: Change in 1930-1939 decadal mean SST anomaly from HadSST.4.1.1.0 to HadSST.4.2.0.0.*

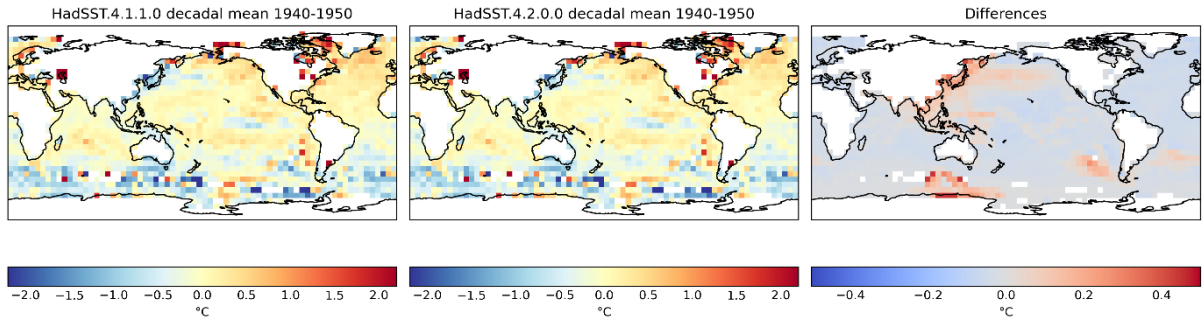

*Figure 10: Change in 1940-1949 decadal mean SST anomaly from HadSST.4.1.1.0 to HadSST.4.2.0.0.*

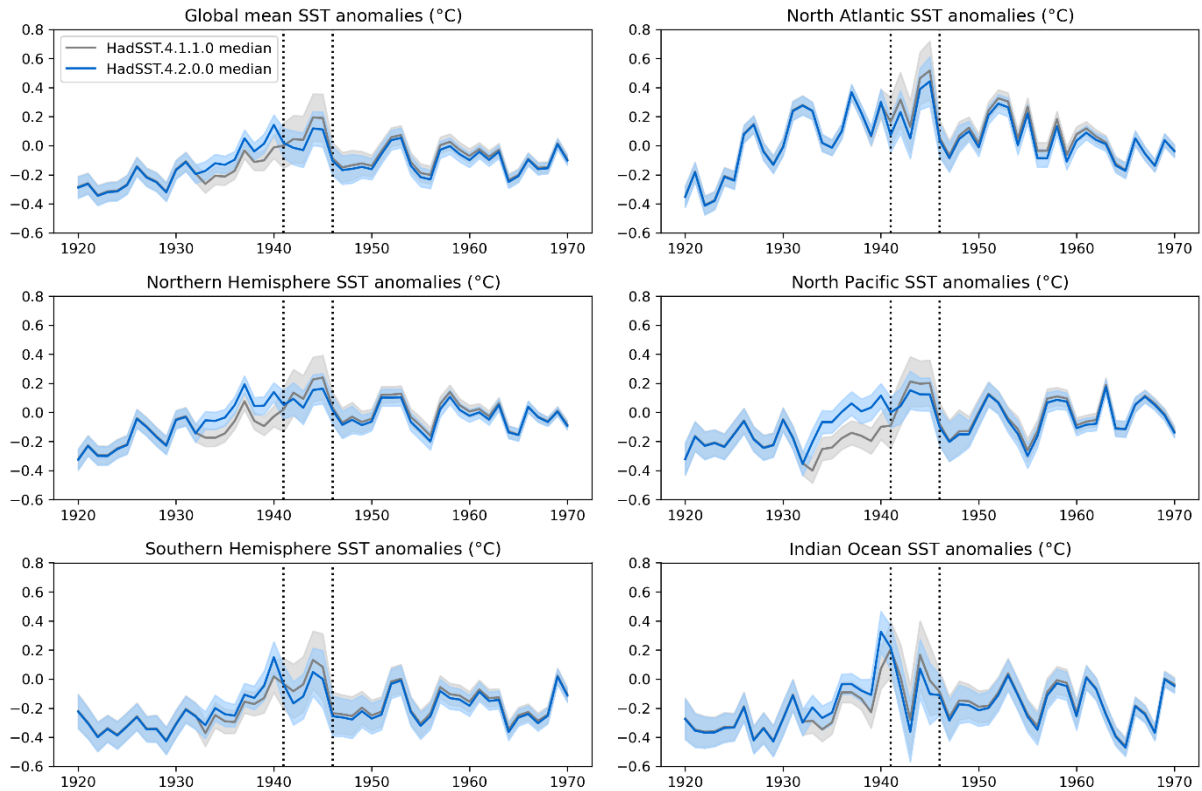

*Figure 11: Annual SST anomaly timeseries and total uncertainties for different regions, focused on the period of influence of decks 118 and 119. Left: globe, Northern Hemisphere and Southern Hemisphere. Right: North Atlantic (0-60N, 70W-0E), North Pacific (0-60N, 130E-110W) and Indian Ocean (50S-20N, 30-110E). Vertical dotted lines mark the beginning and end of the WW2 period as represented in HadSST, from January 1941 to December 1945.*

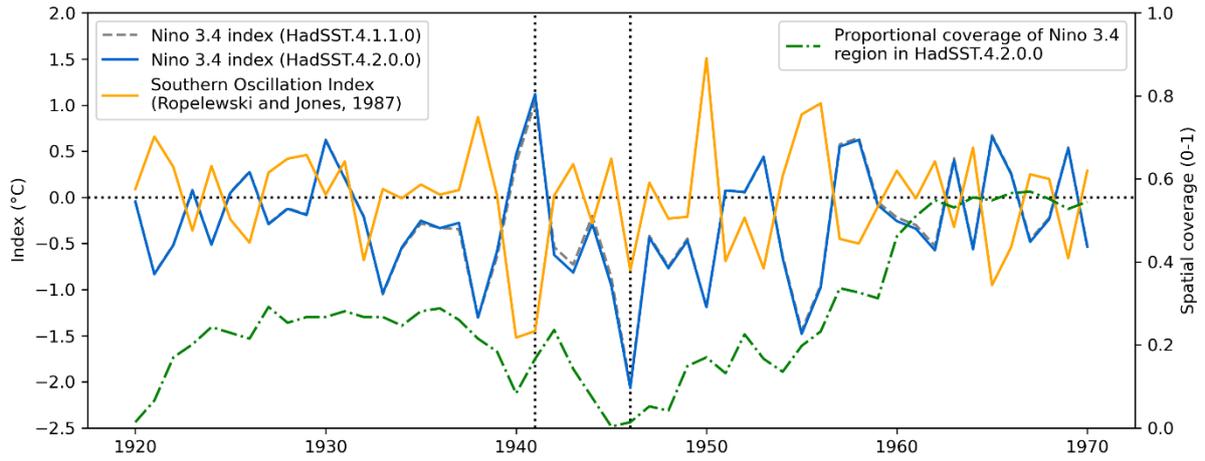

*Figure 12: Niño 3.4 (5S-5N, 120-170W) indices from HadSST.4.2.0.0 and HadSST.4.1.1.0, compared with the Southern Oscillation Index as computed by Ropelewski and Jones (1987) (data accessed July 2025). The green dash-dotted line shows the proportion of grid cells in the Niño 3.4 region (24 5° grid cells in a 2 x 12 arrangement) containing valid data in HadSST.4.2.0.0.*

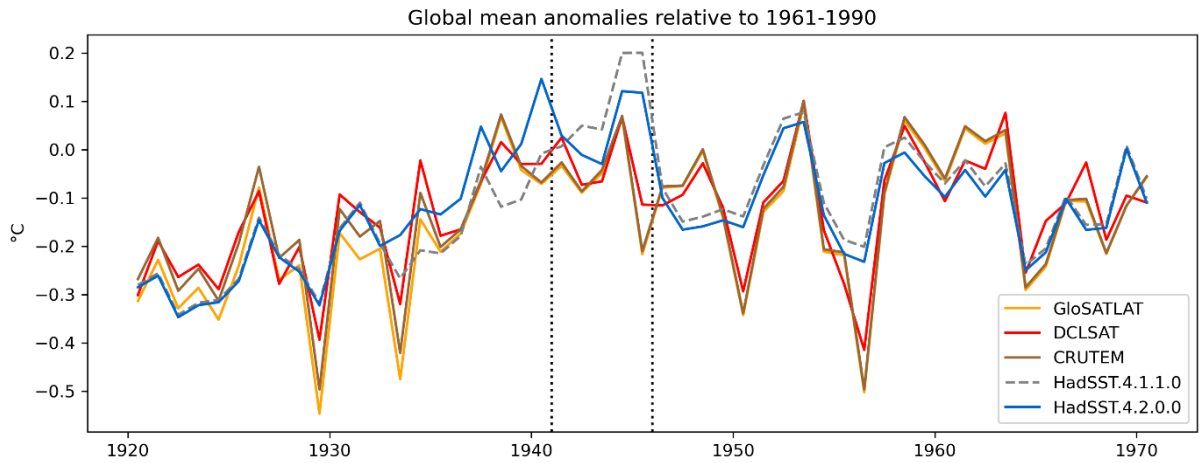

*Figure 13: Comparison of annual global HadSST anomalies with land temperature anomalies from various datasets between 1920 and 1970. Anomalies are defined with respect to the 1961-1990 period in each dataset.*

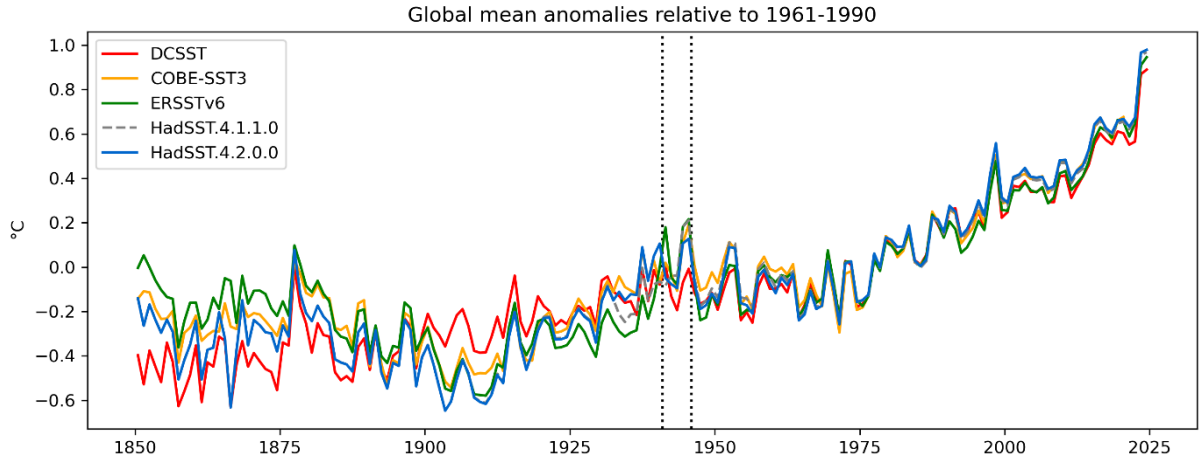

Figure 14: Comparison of annual global anomaly timeseries from five different SST datasets (including two versions of HadSST). Anomalies are defined with respect to the 1961-1990 period in each dataset.

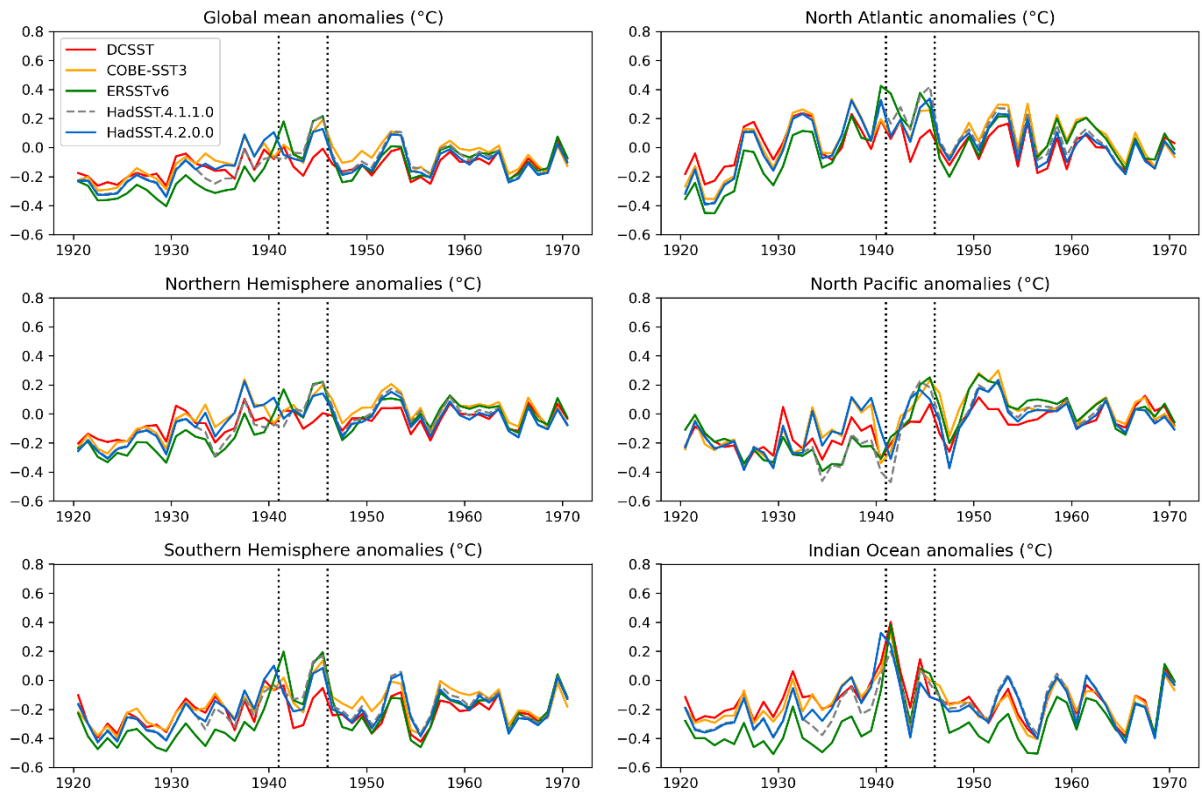

Figure 15: Annual temperature anomaly timeseries with respect to 1961-1990 for selected regions (as defined in Figure 11), for a range of SST datasets.

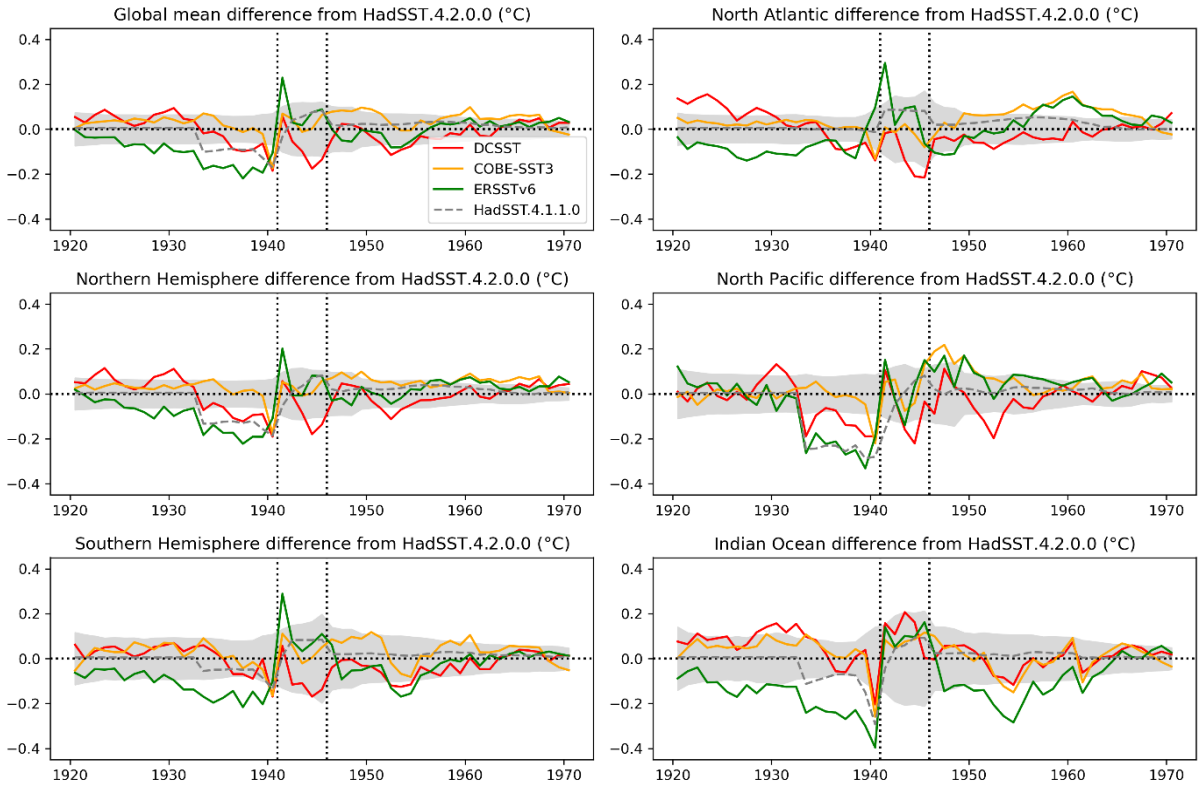

*Figure 16: Annual temperature anomaly differences from HadSST.4.2.0.0 for selected regions (as defined in Figure 11). Grey shaded regions represent the total uncertainty estimate on the regional average from HadSST.4.2.0.0.*

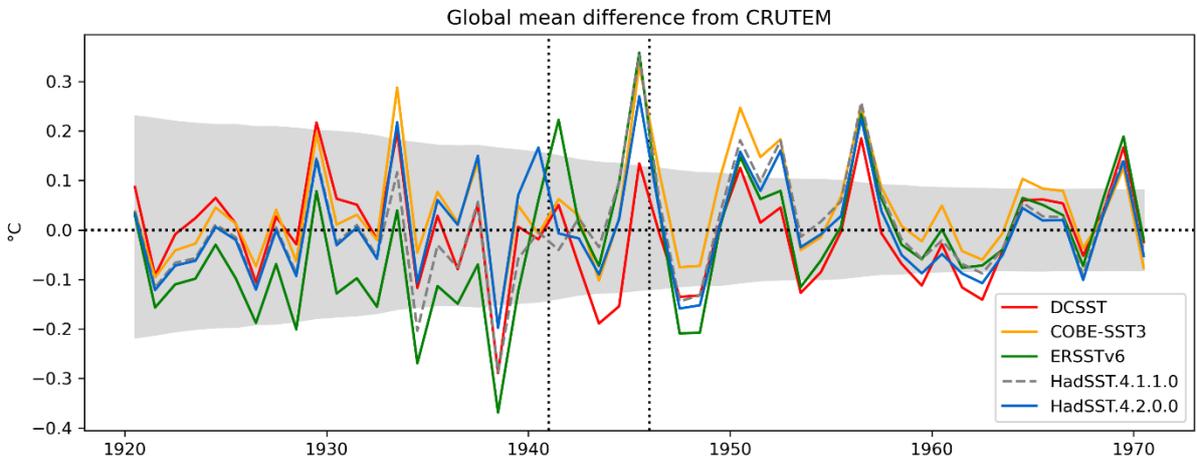

*Figure 17: Comparison of annual global SST anomaly timeseries with an established land temperature timeseries (CRUTEM5, Osborn et al. (2021)) between 1920 and 1970. Grey shaded regions represent the total uncertainty estimate on the CRUTEM timeseries.*